\documentclass{article}
\usepackage{ijcai21}

\usepackage{times}
\usepackage{soul}
\usepackage{url}
\usepackage[hidelinks]{hyperref}
\usepackage[small]{caption}
\usepackage{graphicx}
\usepackage{xcolor}
\usepackage{amsmath}
\usepackage{amsthm}
\usepackage{booktabs}
\usepackage{algorithm}
\usepackage{algorithmic}
\usepackage{amssymb}
\usepackage{subfig}
\usepackage{caption}

\title{A Survey of Machine Learning-Based Physics Event Generation}

\author{
Yasir~Alanazi$^1$\and
N.~Sato$^2$\and
Pawel~Ambrozewicz$^2$\and
Astrid~N.~Hiller~Blin$^2$\and
W.~Melnitchouk$^2$\and\\
Marco~Battaglieri$^2$\and
Tianbo~Liu$^3$\And
Yaohang~Li$^1$
\affiliations
$^1$Department of Computer Science, Old Dominion University, Norfolk, Virginia 23529, USA\\
$^2$Jefferson Lab, Newport News, Virginia 23606, USA\\
$^3$Key Laboratory of Particle Physics and Particle Irradiation (MOE), Institute of Frontier and Interdisciplinary Science, Shandong University, Qingdao, Shandong 266237, China\\
\emails
yalan001@odu.edu,
\{nsato, pawel, ahblin, wmelnitc, battagli\}@jlab.org,
liutb@sdu.edu.cn,
yaohang@cs.odu.edu
}

\begin{document}

\maketitle

\begin{abstract}
Event generators in high-energy nuclear and particle physics play an important role in facilitating studies of particle reactions. 
We survey the state-of-the-art of machine learning (ML) efforts at building physics event generators.
We review ML generative models used in ML-based event generators and their specific challenges, and discuss various approaches of incorporating physics into the ML model designs to overcome these challenges. 
Finally, we explore some open questions related to super-resolution, fidelity, and extrapolation for physics event generation based on ML technology.
\end{abstract}

\section{Introduction}

Accelerators in high-energy nuclear and particle physics have long been the primary source of fundamental knowledge about the nature and interactions of elementary particles and composite hadrons and nuclei.
By accelerating particles to large kinetic energies and colliding them with each other or with stationary targets, the reactions produce a set of outgoing states observed in detectors, and referred to as physics ``events.''
Existing accelerators include the LHC at CERN for proton-proton collisions, CEBAF at Jefferson Lab for polarized electron-nucleon and nucleus scattering, RHIC at Brookhaven National Lab for proton (nucleus)--proton (nucleus) collisions, as well as the future Electron-Ion Collider.
Detecting and analyzing the outgoing particles provides insight into the femtoscale physics underlying these reactions.

Physics event generators, which randomly generate simulated events that mimic those produced in accelerators, play a vital role in facilitating the study of matter.
The event generators have a broad spectrum of physics applications, including estimating the distribution of expected events for the study of interesting physics scenarios, planning and designing new detectors, optimizing the detector performance under experimental constraints, devising strategies to analyze raw data from experiments, and interpreting observed physics phenomena with fundamental theory~\cite{Lehti2010}.

Since the early 1970s, the simulation of physics events has mainly been implemented by Monte Carlo (MC) methods, which transform random numbers into simulated physics events. 
MC event generators (MCEGs)~\cite{Mangano_2005} are constructed by a combination of high-precision data from previous experiments and theoretical inputs.
Commonly used MCEGs include Pythia~\cite{Sjostrand:2007gs}, Herwig~\cite{Bahr:2008pv}, and Sherpa~\cite{Gleisberg:2008ta} for hadronic events; MadGraph~\cite{Alwall_2011} and Whizard~\cite{Kilian_2011} for parton events; GiBUU~\cite{Weil:2012ji} and HIJING~\cite{biro2019introducing} for heavy-ion events; GENIE~\cite{Andreopoulos:2009rq} and NuWro~\cite{juszczak2009running} for neutrino events, as well as specialized event generators such as AcerMC~\cite{Kersevan_2013}, ALPGEN~\cite{Mangano_2003}, and others.

Recently, along with advances in ML, and particularly deep learning, ML-based generative models for physics event generation have received considerable attention.
ML-based event generators (MLEGs) have become an alternative approach to MC simulations of physical processes.
Instead of simulating physics events from first principles as in MCEGs, MLEGs employ a data-driven approach to learn from event samples.
ML generative models, including Generative Adversarial Networks (GANs)~\cite{Goodfellow:2014}, Variational Autoencoders (VAEs)~\cite{vae}, and Normalizing Flows~(NFs) \cite{Kobyzev_2020}, have been adopted to implement~MLEGs.

Compared to MCEGs, MLEGs demonstrate several attractive advantages. 
First, MLEGs can be significantly faster than MCEGs. 
MC simulations of the complete pipeline of particle experiments, including detector effect simulations, often take minutes to generate a single event, even with the help of modern supercomputers~\cite{Buckley:2019wov}. 
In contrast, after proper training MLEGs can produce millions of events per second.
Fast MLEGs can serve as compactified data storage utilities, eliminating the need for maintaining MCEG event repositories~\cite{Chekanov_2015}.
Second, MCEGs rely on theoretical assumptions such as factorization and statistical models, which limit their ability to capture the full range of possible correlations existing in nature between particles' momenta and spins.
On the other hand, MLEGs trained directly on experimental event data are agnostic of theoretical assumptions, and can explore the true underlying probability distributions governing the spectra of particles produced in reactions.

In this paper, we present a survey of ML-based methods for physics event generation. 
Specifically, we review the generative models adopted in state of the art MLEGs and their detector effect simulations. 
In addition to the well-known issues in training a generative model, MLEGs come along with significant physics-related challenges, and we discuss approaches of encoding physics in the ML models to address these challenges. 
Finally, we discuss some open questions about the outlook of MLEG applications, including:
\begin{itemize}
    \item Can MLEGs go beyond the statistical precision of the training event samples?
    \item Can MLEGs faithfully reproduce physics?
    \item Can MLEGs provide new physics insights?
\end{itemize}

\section{Machine Learning-Based Event Generation}

In this section we first review the generative models used in MLEGs, before describing the simulation of detector effects, and discussing the physics-related challenges in MLEGs as well as methods of incorporating physics into ML models to address these.

\begin{table*}[t]
  \centering
\begin{tabular}{ | p{3.5cm} | p{3.6cm} | p{2.8cm} | p{3.6cm} | p{2.2cm} |}
\hline
MLEGs & Data Source & Detector Effect & Reaction/Experiment & ML Model \\
\hline \hline
\cite{hashemi2019lhc} & Pythia8 & DELPHES + pile-up effects & $Z \to \mu^+\mu^-$ & regular GAN \\ 
\hline
\cite{otten2019event} & MadGraph5 aMC@NLO & DELPHES3 & $e^+ e^- \to Z \to l^+ l^-$, $pp \to t\bar t$ & VAE \\ 
\hline
\cite{butter2019GAN} & MadGraph5 aMC@NLO & & $pp \to t \bar{t} \to (bq\bar{q}')(\bar{b}\bar{q}q')$ & MMD-GAN \\
\hline
\cite{DiSipio:2019imz} & MadGraph5, Pythia8 & DELPHES + FASTJET & $2 \to 2$ parton scattering & GAN+CNN \\
\hline
\cite{SHiP:2019gc} & Pythia8 + GEANT4 & & Search for Hidden Particles (SHiP) experiment & regular GAN \\
\hline
\cite{alanazi2020simulation} \cite{cFATGAN}& Pythia8 & & electron-proton scattering & MMD-WGAN-GP, cGAN\\
\hline
\cite{Arjona_Mart_nez_2020} & Pythia8 & DELPHES particle-flow & proton collision & GAN, cGAN\\
\hline
\cite{Gao:2020zvv} & Sherpa & & $pp \to W/Z + n \text{ jets}$ & NF\\
\hline
\cite{howard2021foundations} & MadGraph5 + Pythia8 & DELPHES & $Z \to e^+e^-$ & SWAE\\
\hline
\cite{Choi_2021} & MadGraph5 + Pythia8& DELPHES & $pp \to b \bar{b}\gamma\gamma$ & WGAN-GP\\
\hline
\end{tabular}
\caption{List of existing MLEGs.}
\label{Table:MLEGs}
\end{table*}

\subsection{Generative Models in MLEGs}

\subsubsection{Generative Adversarial Networks}

GANs are the most popularly used generative models in MLEGs. 
The regular GAN event generator is composed of two neural networks: a generator $G$ and a discriminator $D$.
The former is trained to generate fake event samples, and the latter is a binary classifier to distinguish the events of the generated distribution $P_G$ from the true events with distribution $P_T$. $G$ and $D$ are trained under the value function $V(D, G)$
\begin{equation}
\begin{split}
\min_G \max_D V(D, G) 
& = \langle \log D(x) \rangle_{x \sim P_T} \\ 
& + \langle \log (1 - D(\tilde{x})) \rangle_{\tilde{x} \sim P_G}.
\end{split}
\end{equation}
As shown in the original GAN paper~\cite{Goodfellow:2014}, given an optimal discriminator $D^* = P_T(x) / (P_T(x) + P_G(x))$, training $G$ becomes identical to minimizing the Jensen-Shannon divergence (JSD)
\begin{equation}
\min_G V(D^*, G) = -2\log 2 + \text{JSD}(P_G || P_T).
\end{equation}
If JSD becomes $0$, then $P_G=P_T$.

Although GANs have demonstrated success in many applications, training a successful GAN model is known to be notoriously difficult~\cite{Salimans2016ImprovedTF}.
Many GAN models suffer from major problems including mode collapse, non-convergence, model parameter oscillation, instability, vanishing gradient, and overfitting due to unbalanced generator/discriminator combinations. 
Studies \cite{otten2019event,hashemi2019lhc} also reported a less satisfactory performance when a regular GAN is used for event generation.

Several improved GAN architectures have been employed in MLEGs to enhance GAN training:
\begin{itemize}
    \item \textbf{Least Squares GAN (LS-GAN)}: LS-GAN~\cite{mao2017squares} replaces the cross entropy loss function in the discriminator of a regular GAN with a least square term \begin{equation}
    \begin{split}
        \min_D V(D) &= {1 \over 2} \langle(D(x)-b)^2\rangle_{x \sim P_T} \\ 
                    &+ {1 \over 2}\langle(D(G(\tilde{x}))-a)^2\rangle_{\tilde{x} \sim P_G},\\
        \min_G V(G) &= {1 \over 2}\langle(D(G(x))-c)^2\rangle_{x \sim P_G}.
    \end{split}
    \end{equation}
    As a result, by setting $b-a=2$ and $b-c=1$, minimizing the loss function of LS-GAN yields minimizing the Pearson $\chi^2$ divergence.
    The main advantage of LS-GAN is that, by penalizing the samples far away from the decision boundary, the generator is pushed to generate samples closer to the manifold of the true samples.
    \item \textbf{Wasserstein GAN (WGAN)}: WGAN~\cite{arjovsky2017wasserstein} used Wasserstein or Earth-Mover's distance~\cite{Villani2016} to replace JSD in the regular GAN. Under Kantorovich-Rubinstein duality, the Wasserstein distance is defined as
    \begin{equation}
    W(P_G, P_T) = \max_{w \in W} \langle f_w(x) \rangle_{x \sim P_T} - \langle f_w(G(\tilde{x}))\rangle_{\tilde{x} \sim P_G},
    \end{equation}
    where $f$ is a family of $K$-Lipschitz continuous functions, ${f_w}$, parameterized by $w$ in parameter space~$W$.
    Instead of directly telling fake events from the true ones, the discriminator in WGAN is trained to learn a $K$-Lipschitz continuous function to minimize the Wasserstein distance.
    Compared to JSD, Wasserstein distance provides a meaningful and continuous measure of the distance between the event distribution from the generator and the true event distribution, even when they have no overlaps, which helps guide the training of the generator toward the true event distribution and reduce the likeliness of mode collapse.
    
    \item \textbf{Wesserstein GAN Gradient Penalty (WGAN-GP)}: 
    A problem in WGAN is to use weight clipping to maintain $K$-Lipschitz continuity of $f_w$ during training, which still results in unstable training, slow convergence, and gradient vanishing. 
    WGAN-GP~\cite{WGAN_2017} replaces weight clipping with gradient penalty to ensure $K$-Lipschitz continuity and thus further improve WGAN stability. 
    The gradient penalty is calculated as 
    \begin{equation}
        \lambda \langle(|| \nabla_{\hat{x}} f_w(\hat{x})||_2-1)^2\rangle_{\hat{x} \in p _{\hat{x}}},
    \end{equation}
    where parameter $\lambda$ balances the Wasserstein distance and gradient penalty, and $p_{\hat{x}}$ is uniformly sampled along lines between event pairs from $P_T$ and $P_G$.
    
    \item \textbf{Maximum Mean Discrepancy GAN (MMD-GAN)}: MLEGs are particularly concerned about the precise matching between the generated and the true event distributions, where MMD-GAN~\cite{li2017mmd} can be used to enhance the matching precision. MMD-GAN incorporates an MMD term to the generator loss function:
    \begin{equation}
    \begin{split}
    &\text{MMD}^2(P_G, P_T) = \langle k(x, x') \rangle_{x,x' \sim P_G}\\
    &+\langle k(y, y') \rangle_{y,y' \sim P_T}-2\langle k(x, y) \rangle_{x \sim P_G, y \sim P_T},
    \end{split}
    \end{equation}
    where $k(.)$ is a kernel function. MMD typically works well in low dimension; however, the power of MMD degrades with dimension polynomially \cite{Ramdas_2015}.

\end{itemize}

\begin{figure*}[h]
\centering
\subfloat[CLAS detector (downstream view).]{%
  \includegraphics[width=0.27\textwidth]{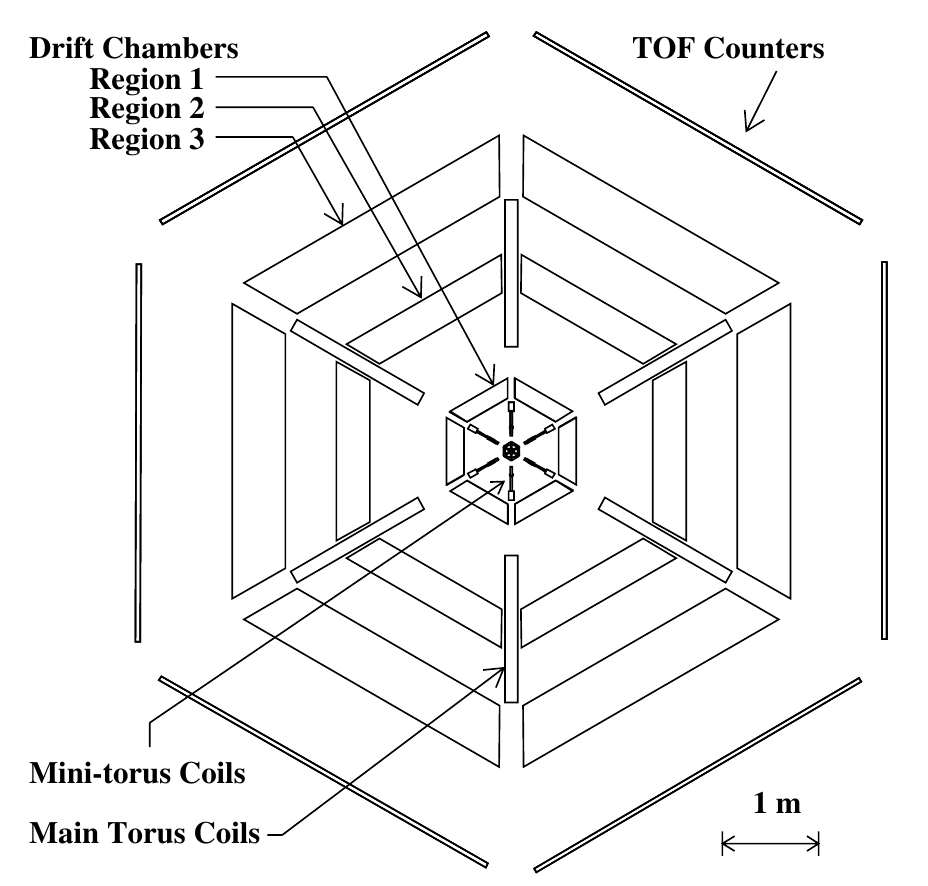}%
  \label{fig:detector}
}
\hfil
\subfloat[$p_x$-$p_y$ plot reflecting detector configuration.]{%
  \includegraphics[width=0.33\textwidth]{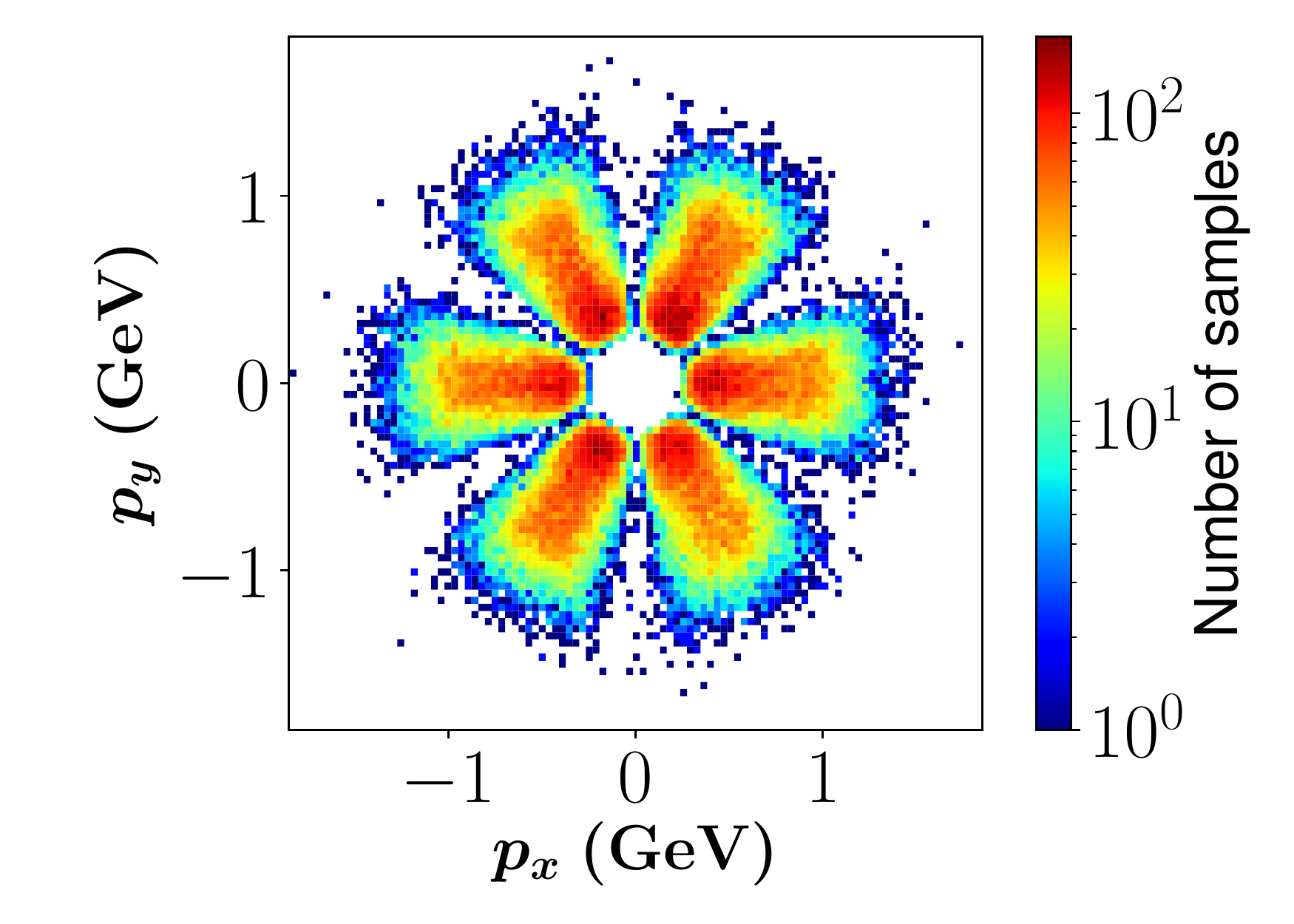}%
  \label{fig:2da}
}\\
\subfloat[$p_x$, $p_y$, $p_z$ distributions, where sharp peaks, deep holes, and steep edges are observed.]{%
  \includegraphics[width=0.9\textwidth]{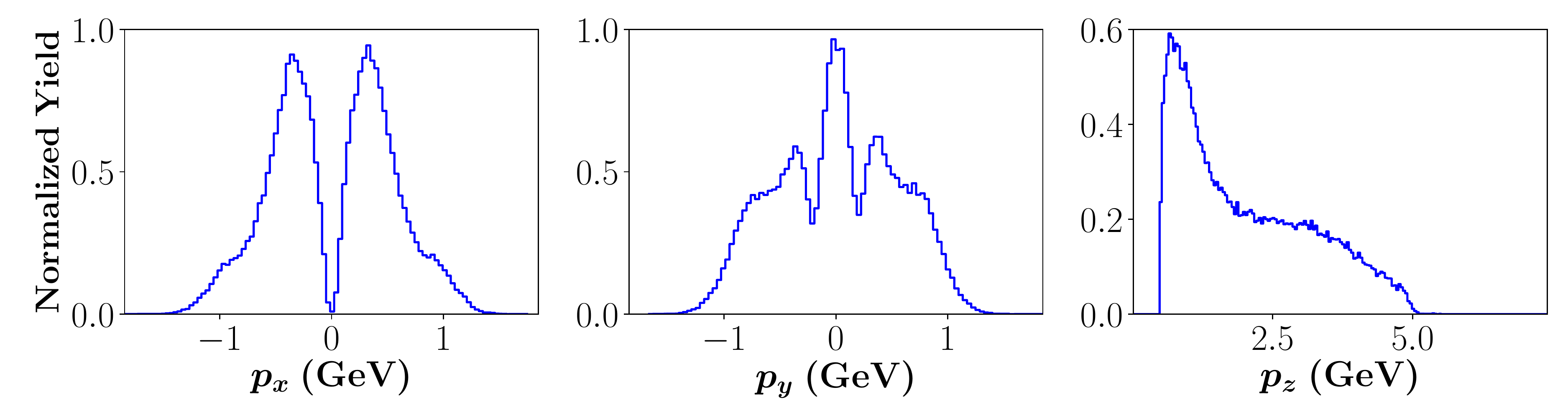}%
  \label{fig:xyz_distribution}
}
\caption{Momentum component distributions of experimental data from an electron scattering experiment with the CLAS detector (a) at Jefferson Lab, with (b) and (c) generated using 75k samples.}
\label{fig:clas}
\end{figure*}

A GAN can also be extended to a conditional GAN (cGAN)~\cite{mirza2014conditional} to be involved in MLEGs to generate events based on an initial reaction condition. 
For example, \cite{cFATGAN} generated electron-proton scattering events conditioned on beam energy. 
\cite{Arjona_Mart_nez_2020} simulated LHC parasitic collisions conditioned on missing transverse energy. 
The condition is injected as an input to the generator along with noise and is then propagated to the discriminator to differentiate fake and true events under the same condition.
The cGAN allows MLEGs to explore events under unseen conditions, either interpolatively or extrapolatively.

\subsubsection{Variational Autoencoder}

A VAE, composed of an encoder network $\Psi$ and a decoder network $\Phi$, is an alternative generative model employed in MLEGs.
In MLEGs using a VAE, $\Psi$ projects the events onto latent variables $z$, and $\Phi$ reconstructs the events from $z$, while $z$ is forced to follow a standard normal distribution.
Then, the loss function of the VAE is motivated by variational inference~\cite{variationalinference} via minimizing the Kullback–Leibler divergence (KLD) between the posterior $p(z|x)$ and the encoded prior distribution $q(z) = \mathcal{N}(0,1)$:
\begin{equation}
    \mathcal{L}_{\rm VAE}=\|x-\Phi(\Psi(x))\|^2+\eta \text{KLD}(q(z) || p(z|x)),
\end{equation}
where the first term is the reconstruction error, the second term computes KLD, and $\eta$ is the harmonic parameter to balance the two.

The VAE can be further improved as a Wasserstein Autoencoder (WAE)~\cite{tolstikhin2018wasserstein} by replacing the KLD term with Wasserstein distance in the loss function, for a similar reason as for the GAN: 
\begin{equation}
    \mathcal{L}_{\rm WAE}=\|x-\Phi(\Psi(x))\|^2+\eta' W(q(z), p(z|x)),
\end{equation}
where $\eta'$ is the harmonic parameter. 
\cite{howard2021foundations} adopted a Sliced Wasserstein Autoencoder (SWAE)~\cite{kolouri2018slicedwasserstein} for their MLEG, using a sliced Wasserstein distance to approximate $W(q(z), p(z|x))$.

\subsubsection{Normalizing Flow}

Without adopting adversarial learning, the NF is another generative model that has been used in MLEGs.
The fundamental idea underlying NFs is the change of variables in probability functions. 
Under some mild conditions, the transformation can lead to complex probability distributions of the transformed variables.
NFs use an invertible mapping (bijection) function $f$, often implemented as a neural network, to transform a distribution of $x \in \mathbb{R}^D$ into $y \in \mathbb{R}^D$. 
The transformed probability density function $q(y)$ becomes
\begin{equation}
    q(y)=p(x)\bigg| \det {{\partial f} \over {\partial x}} \bigg| ^{-1}.
\end{equation}
With a series of mappings $f_1 \dots f_k$, an NF is obtained:
\begin{equation}
    x_k = f_k \circ \dots \circ f_1(x_0), x_0 \sim q_0(x_0).
\end{equation}
The NF is able to transform a simple distribution into a complex multi-modal distribution, and has demonstrated success in collider physics simulations~\cite{Gao:2020zvv}.

\subsubsection{Existing MLEGs}

Table~\ref{Table:MLEGs} lists the existing MLEGs. 
In the literature, GANs, VAEs, NFs, and their various improved architectures have been used to simulate physics events from different reactions and training datasets. 
Both \cite{otten2019event} and \cite{butter2020generative} reported that the LS-GAN yields better performance than other generative models, not only in terms of better precision, but also that in the explored scenarios they were faster.
However, at this point, it is too early to rule out the optimal generative model architecture for general MLEGs, which requires not only computational verification, but also rigorous theoretical justifications. \\

\subsection{Detector Effects}

Any event generator, whether an MCEG or MLEG, that attempts to faithfully reproduce a specific reaction channel must take into account not only the primary interaction at its vertex but also the interactions of the emergent particles with materials and with the devices detecting them.
The former should take into account energy losses, decay, new particle production, as well as multiple scattering effects, while the latter should carefully model detector responses to particles being detected. 
The experimental setup introduces a detection volume, or acceptance, which is usually quite complicated. 
The acceptance covers only a portion of the phase space of the reaction, and has to be modeled employing MC packages, such as GEANT4~\cite{geant4}, DELPHES~\cite{Ovyn2009DELPHESAF}, FLUKA~\cite{Bohlen:2014buj}, or similar.

We classify detector effects into three categories: acceptance, smearing, and misidentification related.
All these effects are mitigated in the MLEGs by using well designed procedures that allow either to remove or to introduce these effects into the synthetic data.
The former is known as ``unfolding'', and the latter as ``folding.''
These procedures usually involve training GANs with additional information introduced by modifying loss functions to improve stability and convergence~\cite{Musella_2018}, using fully conditional GAN (FCGAN), where the conditioning is done on the detector response~\cite{Bellagente_2020}, or employing Wasserstein distance based loss function in a conditional GAN (WGAN) framework~\cite{erdmann2018generating}.

\subsection{Additional Physics-related Challenges}

Compared to many applications employing machine learning generative models to produce images, music, and arts, using MLEGs to simulate events from particle reactions poses new additional physics-related challenges for machine learning:
\renewcommand{\labelenumi}{\roman{enumi}}
\begin{enumerate}
\item Events generated by MLEGs should not violate physical laws, such as energy and momentum conservation;
\item MLEGs for generating particle physics events are required to model the distributions of event features and their correlations sufficiently precisely for the nature of particle reactions to be correctly replicated;
\item The distributions of events exhibit natural, physics driven patterns, such as discrete attributes, prominent and narrow peaks, or symmetric behavior of certain physical quantities. On top of that they also exhibit artificial, detector-related, patterns, such as acceptance induced holes and gaps, and efficiency based regions of lower particle occupancy, which complicate MLEGs; 
\item The outgoing particles, with increasing incident energy, will yield increased dimensionality of the emergent products.
\end{enumerate}

It is important to note that all existing MLEGs listed in Table~\ref{Table:MLEGs} are trained using simulated data generated from MCEGs, such as Pythia, MadGraph, and Sherpa. When learning from real experimental data, it adds an additional level of complications to the MLEGs. 
Figure~\ref{fig:clas} shows the momenta plots of experimental data from an electron scattering experiment with the CEBAF Large Acceptance Spectrometer (CLAS) detector at Jefferson Lab. 
Due to the configuration of superconducting coils of the torus magnet, spaced by angles of $60^{\circ}$, the detector packages~\cite{ADAMS2001414} shown in Fig.~\ref{fig:detector} are accordingly divided into six sectors, so that events which fall into the coils are not detected, leaving six gaps in any particle transverse momentum components plot, $p_x$ and $p_y$, as in Fig.~\ref{fig:2da}. 
As a result, the 1D plots of $p_x$, $p_y$, and the remaining longitudinal component, $p_z$, as shown in Fig.~\ref{fig:xyz_distribution}, yield spikes, deep holes, and sharp edges, which pose difficulty for MLEGs to precisely learn their inherent physical laws as well as the detector patterns.

\subsection{Incorporating Physics into ML Models}
\label{physicsML}

To address the above physics-related challenges, an important approach is to incorporate physical laws into the generative models.
Appropriately encoding physical laws into the models can reduce the degrees of freedom of the problem and improve the performance of MLEGs.

\begin{center}
  \includegraphics[width=0.9\linewidth]{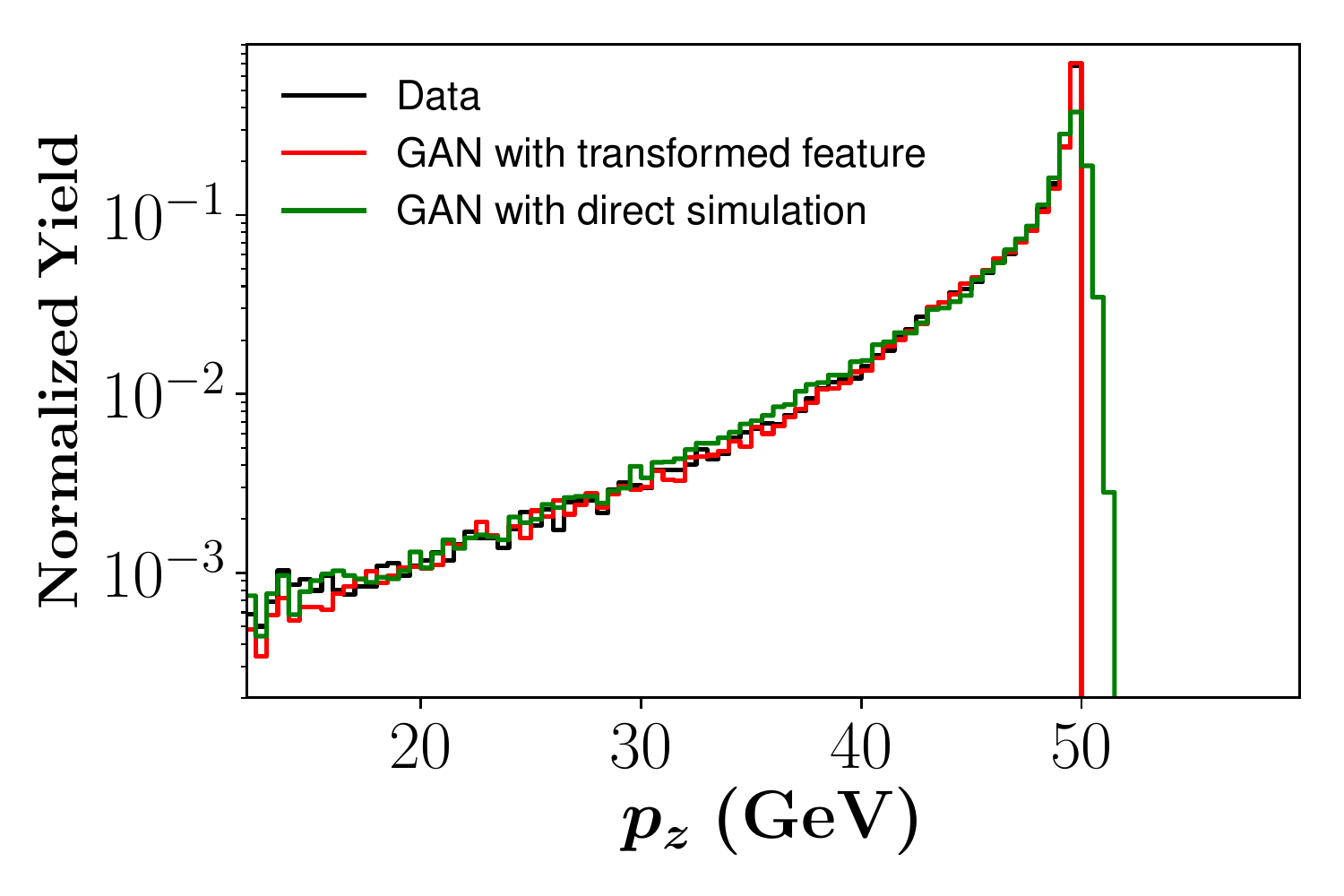}
  \captionof{figure}{Comparison of the $p_z$ distributions generated by the GAN with the transformed features (red), and the direct simulation GAN (green), and the true distribution from Pythia (black).}
  \label{fig:pz}
\end{center}

One way to incorporate a physical law into the generative models is via feature engineering. 
When simulating inclusive scattering of electrons, \cite{alanazi2020simulation} found that a direct simulation GAN generates unphysical events that violate energy conservation. 
As shown in Fig.~\ref{fig:pz}, a sharp edge in the particle energy $E$ distribution arises from energy conservation, which restricts $E$ to be less than the input beam energy, $E_{\rm b}$.
This sharp edge is difficult to learn for the inclusive GAN, whose output is the electron momentum 3-vector $(p_x, p_y, p_z)$, as unphysical events can be generated with $E > E_{\rm b}$, which the discriminator is not sensitive enough to differentiate from the eligible physics events, particularly when $E_{\rm b} - E$ is small.

To address this problem, a transformation ${\cal T}(p_z) = \log(E_{\rm b} - p_z)$ is applied to replace $p_z$ as the output variable of the generator, which avoids the production of unphysical particles. 
As shown in Fig.~2, the transformation ${\cal T}(p_z)$ improves the sensitivity of the discriminator, yielding a significantly better match of the $p_z$ distribution with data. 
In another example, \cite{hashemi2019lhc} took into account the symmetries of the process $Z \to \mu^+\mu^-$ and pre-processed the event samples so that the azimuthal angle of the leading charged lepton is always zero, which resulted in a substantial improvement in terms of agreement with the testing samples.

Another approach to incorporate physics into generative models is to make the latent variables physically meaningful. 
Typically, the noise fed to GAN or the latent variables in VAE follow certain well-known, easy-to-generate distributions such as Gaussian or uniform without physics meanings. 
\cite{howard2021foundations} expressed the latent distributions in SWAE with quantum field theory and thus sampling the latent space became efficient and was able to infer physics.

\section{Open Questions}

Compared to MCEGs that have been developed for over 50 years, MLEGs are still in their infant stage, bringing a lot of anticipation as well as many challenges and questions currently without clear answers.
In this section, we discuss three open questions regarding the applications of MLEGs.

\subsection{Can MLEGs Display Super-Resolution?}

One of the very attractive properties of generative models is super-resolution~\cite{SRGAN}, or generating samples going beyond the resolution of its training samples. 
Correspondingly, in physics event generation a question that has been debated in literature is whether the generated events can add statistics beyond that of the training sample or not. 
That is, can the MLEG generate data that does not only reproduce the examples seen in the training data, but produces additional, diverse, and realistic samples that are more useful for downstream tasks than the original training data?

It has been claimed in~\cite{Matchev:2020tbw} that, since the network does not add any physics knowledge, one can only achieve as much statistical precision as in the training sample.
The main reason for this statement was that an MLEG does not learn to mimic the true event generator of the training sample, but rather the data of the sample it was trained on. 
This would imply that one cannot improve the model or parameter discrimination by increasing statistics with the MLEG.
In addition, the statistical uncertainty of the training samples would enter the MLEG as systematic uncertainty, thus creating an even more stringent overall uncertainty than in the original data. 
It is stated, nevertheless, that the MLEG could still offer a better relation between accuracy and computational and storage resources than MCEGs or real experimental data.

The findings of~\cite{butter2020ganplifying} show empirically that the claim of~\cite{Matchev:2020tbw} is neither well-founded nor fulfilled, thus declaring MLEG as a promising venue for the amplification of training statistics.
Moreover, this work quantifies the extent to which the events can be amplified before being limited by the statistics of the training samples.
The argument in favor of augmentation feasibility relies on the fact that MLEGs are powerful interpolation tools even in high-dimensional spaces.
Despite being model-agnostic interpolators, the interpolation functions do fulfill basic properties such as smoothness, and therefore can add to the discrete data sets in a reliable fashion, by enabling denser binning, or higher resolution. 
In fact, it was shown that an MLEG can achieve the same precision as the minimal precision of a fit with a known functional form, saturating when the number of generated events becomes orders of magnitude larger than the size of the original training set. 
While this means that the MLEG cannot outperform the precision of a functional fit, it reassures us that in those cases where the functional form of the fitting curve is unknown (which represents most realistic physics scenarios), the MLEG becomes a powerful tool for precision augmentation.

\subsection{Can MLEGs Faithfully Reproduce Physics?}

Whether MLEGs can fully represent the underlying physics of a reaction and faithfully reproduce physical events is critical to many applications where MLEGs are proposed.
In the existing MLEGs listed in Table~\ref{Table:MLEGs}, the agreement between the events generated by MLEGs and the sample events is mostly measured in $1$D using $\chi^2$ or other statistical metrics.
In practice, one is often interested in the correlations among event feature distributions. 
Another issue is related to the rare events occurring in the reaction, which are often precisely those of significant physics interest.
Such rare events pose a difficult challenge to MLEGs, however, which often bias to more frequent events during their training.

The methods of incorporating physics into MLEGs, as described in Sec.~\ref{physicsML}, help improve the precision of MLEGs.
Augmenting the generated features of MLEGs to other important physics properties of interest, as described in~\cite{alanazi2020aibased}, also increases the sensitivity of the discriminator in the GAN event generator, and thus enhances the quality of the generated events.
Nevertheless, at this point, there is lack of a comprehensive evaluation framework to thoroughly evaluate the quality of MLEG events in comparison with those from MCEGs or from experiment, particularly in quantifying the correlation among event features with physics meaning, as well as measuring the quality of the rare events.

\subsection{Can MLEGs Provide New Physics Insights?}

Can an MLEG go beyond the manifold of its training event data and bring physical insight into regions without any training (experimental) data? 
We start the discussion of this question from the extrapolation capability of neural networks. 
A major drawback of a neural network is its difficulty with extrapolation.
The theoretical explanation is rooted in the ``universal approximation property'' of a feed-forward neural network~\cite{SCARSELLI199815}, {\it i.e.}, a neural network can approximate any continuous function and many discontinuous functions by adjusting its parameters with respect to its training samples. 
Therefore, for the space outside of the range of the training samples, the output of a neural network is not reliable. 
MLEGs trained using GANs, VAEs or NFs are fundamentally neural networks, which thus inherit the extrapolation problem. 
\cite{cFATGAN} showed that an MLEG based on cGAN yields good agreement for interpolating events between training beam energy levels, but not as good agreement for extrapolating events beyond training beam energy levels, particularly for those related to beam energies further from the training beam energies.

There are two potential ways to extend MLEGs to generate correct events in unknown regions. 
One way is to apply regularization, forcing the generator to adopt a simple model to limit the degrees of freedom of its neural network and avoid overfitting. 
More specifically, incorporating known physical laws into the regularizers helps generalize the neural networks and reduce their variation to explore the unknown physical laws and theories. 
The other way is to generate artificial data samples within the unknown region by either physics theory or simulation to correct the behavior of MLEGs.
Both approaches can also be combined to allow MLEGs, at least at some extent, to extrapolate.

\section{Conclusion and Outlook}

Along with advances in ML methods, MLEGs are emerging as an alternative approach to MCEGs for generating simulated physical events that mimick those produced in high-energy physics accelerators.
Compared to MCEGs, MLEGs demonstrate clear advantages, including fast event generation and being agnostic of theoretical assumptions.

The development of MLEGs is still in its relatively early stages.
In this survey paper, we review ML generative models used in existing MLEGs, including GANs, VAEs, NFs, and their enhanced architectures. 
Some studies reported that LS-GAN yields better event quality over other generative architectures, but these lack theoretical justifications.
MLEGs, as practical tools to simulate physical events, pose additional challenges related to physics, and we review methods of incorporating physics into MLEGs to address these challenges.
We further explore the open questions on the capability of MLEGs on super-resolution, faithful reproduction of physics, and extrapolation.
Answers to these open questions will have significant impact on the applications of MLEGs.

It is important to note that MLEGs are not likely to replace MCEGs, which are used to verify the underlying theory when compared with experimental data. 
MLEGs, on the other hand, can serve the purpose of remedying the statistical weakness of MCEGs, particularly if the their super-resolution capability is confirmed. 
If MLEGs are extended with extrapolation capability, they may interest broad applications to bring in new physics insights. 
When MLEGs' faithfulness of reproducing physics is well-justified, MLEGs can also be used as a compactified data storage utility to efficiently store and regenerate physical events.

Unlike many generative model applications, such as producing sharp looking images or fancy objects, where the distribution agreement between the generated samples and the truths is often not strictly enforced, the general requirements underlying MLEGs are to precisely reproduce a specific target distribution. 
The generative models developed in MLEGs, which incorporate domain knowledge into the machine learning algorithms to faithfully generate samples mimicking complex target distributions, have the potential to be applied to broader applications, such as bioinformatics \cite{hicGAN} and cosmology \cite{Li_2021}.

\section*{Acknowledgements}

We thank the CLAS Collaboration, particularly F.~Hauenstein, for assistance with extracting the CLAS electron scattering data, and J.~Qiu for helpful discussions.
This work was supported by the LDRD project No.~LDRD19-13, No.~LDRD20-18, and No.~LDRD21-22.

\bibliographystyle{named}
\bibliography{ijcai21}

\end{document}